\documentclass[prb,aps,twocolumn,showpacs,ams]{revtex4-1}
\usepackage{graphicx}
\usepackage{dcolumn}
\usepackage{latexsym}
\usepackage{color}

\begin{document}


\title{
Site-selective NMR for odd-frequency Cooper pairs around vortex\\
 in chiral $p$-wave superconductors
}



\author{Kenta K. Tanaka} 
\email[]{ktanaka@mp.okayama-u.ac.jp}
\affiliation{
Department of Physics, Okayama University, Okayama 700-8530, JAPAN}

\author{Masanori Ichioka}
\email[]{ichioka@cc.okayama-u.ac.jp}
\affiliation{
Department of Physics, Okayama University, Okayama 700-8530, JAPAN}

\author{Seiichiro Onari} 
\affiliation{
Department of Physics, Okayama University, Okayama 700-8530, JAPAN}


\date{\today}

\begin{abstract}
In order to identify the pairing symmetry with chirality,
we study site-selective NMR in chiral $p$-wave superconductors. 
We calculate local nuclear relaxation rate $T_1^{-1}$ 
in the vortex lattice state by Eilenberger theory, 
including the applied magnetic field dependence. 
We find that $T_1^{-1}$ 
in the NMR resonance line shape  
is different between two chiral states $p_{\pm}(=p_x{\pm}ip_y)$, 
depending on whether the chirality is parallel or anti-parallel to 
the vorticity.
Anomalous suppression of $T_1^{-1}$ occurs around the vortex core
in the chiral $p_-$-wave
due to the negative coherence term 
coming from the odd-frequency $s$-wave Cooper pair induced 
around the vortex
with Majorana state.  

\end{abstract}



\maketitle

\section{Introduction}
In the study of unconventional superconductors,
it is most important to identify the spin and orbital symmetry of the Cooper pairs
since it is tightly related to the mechanism of superconductivity.
The pairing symmetry of the ruthenate superconductor 
$\rm{Sr_2RuO_4}$ is suggested to be 
chiral $p_\pm$-wave~\cite{MackenzieMaeno,Sr2RuO4-kappa}, 
where Cooper pairs have angular momentum $L_z =\pm 1$ for 
$p_\pm=p_x \pm i p_y$. 
For experimental evidence,
the spin triplet paring is supported by the Knight shift measurement~\cite{Sr2RuO4-NMR}
and the broken time-reversal symmetry coming from the chiral pair was 
observed by ${\rm \mu SR}$~\cite{Sr2RuO4-muSR} 
and polar Kerr effect~\cite{Xia} measurements.
However, any experiment to identify the direction of the chirality, i.e., $p_+$ or $p_-$ in $\rm{Sr_2RuO_4}$ is not yet realized,
since the ${\rm \mu SR}$ and the polar Kerr effect measurements can only detect the existence of chirality ($L_z=0$ or ${\neq}0$).

The spatially resolved NMR measurement~\cite{ssNMR1, ssNMR2, ssNMR3, ssNMR4} called site-selective NMR
can detect local electronic states related to the pairing symmetry 
in the vortex lattice state
by selectively observing 
the resonance field dependence 
of the nuclear relaxation rate $T_1^{-1}$ in the NMR resonance line shape. 
This measurement is complementary method to the scanning tunneling microscopy measurement,
since the NMR measurement is free from the material surface condition.
From our previous studies for site-selective NMR~\cite{Takigawa, K.Tanaka2},  
local $(T_1T)^{-1}$ in the vortex lattice state is determined by 
local density of states (DOS) of electrons 
in the $s$- and $d_{x^2-y^2}$-wave superconductors.
As for the chiral $p$-wave superconductor, 
previous theories suggest that
the temperature $T$-dependence of $T_1^{-1}$ is different between $p_+$ and $p_-$ states 
at the vortex center~\cite{Takigawa-chiral, Hayashi-Kato, Kato-NMR}. 
This chirality-dependence is caused by the interaction between the chirality and vorticity,
depending on whether the chirality $L_z(=\pm 1)$ is parallel or 
anti-parallel to the vorticity $W(=1)$ 
in the vortex state of chiral $p$-wave superconductors
~\cite{GL-chiral, Ichioka-chiral, meta-chiral, Tanuma}.

Recently, the chiral $p$-wave superconductors have been attracting much attention 
as a topological superconductor, 
since it has non-trivial topological properties.
In this superconductor, 
topological defects such as vortex or surface induce Majorana fermions~\cite{ReadGreen, Ivanov,Mizushima}.
Majorana fermions give rise to anomalous electric states such as Majorana zero mode 
and non-Abelian statistics of the vortices~\cite{Ivanov}. 
In addition, the vortex state of chiral $p$-wave superconductors also induces 
odd-frequency Cooper pairs~\cite{Tanuma, Y.Tanaka, Daino}. 
In particular, the odd-frequency $s$-wave Cooper pair in the vortex state of chiral $p$-wave superconductors
is related to the Majorana fermion~\cite{Daino}.

The purpose of this paper is that we investigate
the method to identify the pairing symmetry with chirality 
by the site-selective NMR measurement.
In chiral $p$-wave superconductors,
it is significant to prove topological numbers $L_z$ and $W$ as well as local DOS.
From this view point, 
we study the chirality-dependence of local $T_1^{-1}({\bf{r}})$ 
in the resonance field dependence
in the vortex lattice state. 
We especially focus on anomalous suppression of $T_1^{-1}$ around the vortex core in the chiral $p_-$-wave.
Further, 
we will discuss reasons for the anomalous suppression of $T_1^{-1}$ 
in the relation to 
odd-frequency Cooper pairs induced around the vortex 
with Majorana state.  

This paper is organized as follows.
After the introduction, 
we explain our formulation of Eilenberger theory for the vortex lattice state,
and calculation method for $T_1^{-1}$ in Sec. II.
In Sec. III, we study the temperature, spatial and resonance field dependence of 
local $T_1^{-1}({\bf{r}})$ in the vortex lattice state.
In Sec. IV, we discuss the reasons for the anomalous suppression of $T_1^{-1}$.
The last section is devoted to summary.

\section{Formulation}
We calculate the spatial structure of the vortex lattice state 
by quasiclassical Eilenberger theory
~\cite{Ichioka-chiral, K.Tanaka1, K.Tanaka2}.
The quasiclassical theory is valid when the atomic scale is enough 
small compared to the superconducting coherence length $\xi$.
For many superconductors including $\rm{Sr_2RuO_4}$, this quasiclassical condition is well satisfied~\cite{MackenzieMaeno,Sr2RuO4-kappa}.
Moreover, since our calculations are performed in the vortex lattice state,
distributions of local $T_1^{-1}$ and the resonance field are 
quantitatively obtained as a function of temperature and applied field.
Therefore, our calculation method 
is powerful and reliable tool dealing with the inhomogeneous spatial structure of superconducting properties.

As a simple model of $\rm{Sr_2RuO_4}$, we consider the chiral $p$-wave pairing 
on the cylindrical Fermi surface, 
${\bf k}=(k_x,k_y)=k_{\rm F}(\cos\theta_k,\sin\theta_k) $, 
and the Fermi velocity 
${\bf v}_{\rm F}=v_{\rm F0} {\bf k}/k_{\rm F}$. 
Quasiclassical Green's functions
$g({\rm i}\omega_n,{\bf k},{\bf r})$, 
$f({\rm i}\omega_n,{\bf k},{\bf r})$,
$f^\dagger({\rm i}\omega_n,{\bf k},{\bf r})$ 
are calculated 
by solving 
Eilenberger equation
\begin{eqnarray} &&
\left\{ \omega_n 
+{\bf v} \cdot\left(\nabla+{\rm i}{\bf A}({\bf r}) \right)\right\} f 
= \tilde\Delta({\bf r},{\bf k}) g, 
\nonumber
\\ && 
\left\{ \omega_n 
-{\bf v} \cdot\left( \nabla-{\rm i}{\bf A}({\bf r}) \right)\right\} f^\dagger
= {\tilde\Delta}^\ast({\bf r},{\bf k}) g  , \quad 
\label{eq:Eil}
\end{eqnarray} 
where $g=(1-ff^\dagger)^{1/2}$, 
and
${\bf v}={\bf v}_{\rm F}/v_{{\rm F}0}$.
The order parameter is 
$\tilde\Delta({\bf r},{\bf k})
=\Delta_+({\bf r}) \phi_{p+}({\bf k}) + \Delta_-({\bf r}) \phi_{p-}({\bf k})$
with the pairing function 
$\phi_{p \pm}({\bf k})=(k_x {\pm} ik_y)/k_{\rm F}={\rm e}^{\pm i \theta_k}$ 
for the chiral $p_{\pm}$-wave. 
${\bf r}$ is the center-of-mass coordinate of the pair. 
When magnetic fields are applied along the $z$ axis, 
the vector potential is given by   
${\bf A}({\bf r})=\frac{1}{2} {\bf H} \times {\bf r}
 + {\bf a}({\bf r})$ in the symmetric gauge, 
where ${\bf H}=(0,0,H)$ is a uniform flux density, 
and ${\bf a}({\bf r})$ is related to the internal field 
${\bf B}({\bf r})=(0,0,B({\bf r})) 
={\bf H}+\nabla\times {\bf a}({\bf r})$. 
We have scaled temperature, length, and magnetic field
in unit of $T_{c_0}$, $\xi_0$, and $B_0$, 
where $\xi_0=\hbar v_{\rm F0}/2 \pi k_{\rm B} T_{c_0}$, 
$B_0=\phi_0 /2 \pi \xi_0^2$ with the flux quantum $\phi_0$, respectively.  
$T_{\rm{c_0}}$ is transition temperature at a zero field. 
The energy $E$, pair potential $\Delta$ and Matsubara frequency $\omega_n$ 
are in unit of $\pi k_{\rm B} T_{\rm c_0}$. 
In the following, we set $\hbar=k_{\rm B}=1$. 

To determine $\Delta_\pm({\bf r})$ and 
the quasiclassical Green's functions selfconsistently, 
we calculate $\Delta_\pm({\bf r})$ by the gap equation
\begin{eqnarray}
\Delta_\pm({\bf r})
= g_0N_0 T \sum_{0 < \omega_n \le \omega_{\rm cut}} 
 \left\langle \phi_{p \pm}^\ast({\bf k}) \left( 
    f +{f^\dagger}^\ast \right) \right\rangle_{\bf k} , 
\label{eq:scD} 
\end{eqnarray} 
where 
$(g_0N_0)^{-1}=  \ln T +2 T\sum_{0 < \omega_n \le \omega_{\rm cut}}\omega_n^{-1} $, 
and we use $\omega_{\rm cut}=20 k_{\rm B}T_{\rm c_0}$. 
$\langle \cdots \rangle_{\bf k}$ indicates the Fermi surface average.
For the selfconsistent calculation of the vector potential 
for the internal field $B({\bf r})$,  
we use the relation
\begin{eqnarray}
\nabla\times \left( \nabla \times {\bf A} \right) 
=- 2T{\kappa}^{-2}  \sum_{0 < \omega_n} 
 \left\langle {\bf v} {\rm Im} \{ g  \} 
 \right\rangle_{\bf k}. 
\label{eq:vec} 
\end{eqnarray}
In our calculations, 
we use the Ginzburg-Landau parameter~\cite{K.Tanaka1, K.Tanaka2, kappa_Miranovic} $\kappa=2.7$ appropriate to 
${\rm Sr_2RuO_4}$~\cite{MackenzieMaeno,Sr2RuO4-kappa}.


We iterate calculations of Eqs. (\ref{eq:Eil})-(\ref{eq:vec})
in Matsubara frequency $\omega_n$ in the square vortex 
lattice~\cite{Sr2RuO4-square1}, 
until we obtain the selfconsistent results of 
${\bf{A}}({\bf{r}})$, ${\Delta}({\bf{r}})$ and 
the quasiclassical Green's functions. 
We consider two states of $p_\pm$. 
In the $p_{+}$ state, where chirality and vorticity are parallel, 
$\Delta_{+}({\bf r})$ is main component and $\Delta_{-}({\bf r})$ is 
induced around vortices. 
In the $p_-$ state where $\Delta_-({\bf r})$ is main component, 
chirality and vorticity are anti-parallel. 
The studies of phase diagram for thermodynamically stable states have been already reported
in Refs. \cite{GL-chiral, Ichioka-chiral}. 
According to these previous studies, 
the $p_-$ state has a lower free energy than the metastable $p_+$ state
in all $T$-$H$ range except for $H=0$.
At the $H=0$, the chiral $p_{\pm}$ states are degenerate in free energy. 
We study not only the stable $p_-$ state case but also the metastable $p_+$ state case.
From our calculation results, the upper critical field is $H_{c2}/B_0 = 0.84$ at $T/T_{c_0} = 0.5$ 
for the $p_-$ state. 
The $p_+$ state is unstable 
at $H/B_0 > 0.31$ at $T/T_{c_0}= 0.5$, and changes to the $p_-$ state.


Next, using the selfconsistently obtained 
${\bf{A}}({\bf{r}})$ and ${\Delta}({\bf{r}})$, 
we calculate quasiclassical Green's functions 
in real energy $E \pm {\rm i}\eta$ instead of ${\rm i}\omega_n$. 
Since we consider the clean case
with long lifetime of quasiparticle, 
we use enough small $\eta(=0.01)$,
maintaining the accuracy of numerical calculation.
We solve  Eilenberger equation (\ref{eq:Eil}) with 
${\rm i}\omega_n \rightarrow E \pm {\rm i}\eta$ 
to obtain $g(E\pm{\rm i}\eta,{\bf k},{\bf r})$, 
$f(E\pm{\rm i}\eta,{\bf k},{\bf r})$, 
$f^\dagger(E\pm{\rm i}\eta,{\bf k},{\bf r})$. 
The local DOS $N(E,{\bf r})$ is given by 
$N(E,{\bf r})
=\langle {\rm Re} \{ g(E+{\rm i}\eta,{\bf k},{\bf r}) \}\rangle_{\bf k}$. 

Based on the linear response theory, 
from the obtained quasiclassical Green's functions,   
the nuclear relaxation rate $T_1^{-1}$ 
is calculated as~\cite{Hayashi-Kato, K.Tanaka2} 
\begin{eqnarray} && 
\frac{(T_{1}(T)T)^{-1}}{(T_{1}(T_{\rm c}) T_{\rm c})^{-1}}
=\frac{(T_{1gg}(T)T)^{-1} + (T_{1ff}(T)T)^{-1}}
      {(T_{1}(T_{\rm c}) T_{\rm c})^{-1}}
\nonumber \\ && 
{\hspace{20.5mm}}={\int^{\infty}_{-\infty}}
\frac{W_{gg}({E},{\bf r})+W_{ff}({E},{\bf r})}{4T \cosh^{2}(E /2T)} {\rm d}E, 
\qquad 
\label{eq:T1} 
\end{eqnarray}
where 
\begin{eqnarray} && 
W_{gg}({E},{\bf r})
= \langle a_{{\downarrow}{\downarrow}}^{22}({E},{\bf k},{\bf r})
  \rangle_{\bf{k}}
  \langle a_{{\uparrow}{\uparrow}}^{11}({-E},{\bf k},{\bf r})
  \rangle_{\bf{k}}, 
\nonumber 
\\ && 
W_{ff}({E},{\bf r})
=-\langle a_{{\downarrow}{\uparrow}}^{21}({E},{\bf k},{\bf r})
  \rangle_{\bf{k}}
  \langle a_{{\uparrow}{\downarrow}}^{12}({-E},{\bf k},{\bf r})
  \rangle_{\bf{k}} 
\label{eq:Wff}
\end{eqnarray}
with 
\begin{eqnarray} && 
a_{{\uparrow}{\uparrow}}^{11}({E},{\bf k},{\bf r}) 
=\frac{1}{2} \left[ 
  g({E}+i{\eta},{\bf k},{\bf r})
 -g({E}-i{\eta},{\bf k},{\bf r}) \right],
\nonumber
\\ && 
a_{{\downarrow}{\downarrow}}^{22}({E},{\bf k},{\bf r}) 
=\frac{1}{2} \left[ 
  \bar{g}({E}+i{\eta},{\bf k},{\bf r})
 -\bar{g}({E}-i{\eta},{\bf k},{\bf r}) \right],
\nonumber 
\\ && 
a_{{\uparrow}{\downarrow}}^{12}({E},{\bf k},{\bf r}) 
=\frac{i}{2} \left[ 
  f({E}+i{\eta},{\bf k},{\bf r})
 -f({E}-i{\eta},{\bf k},{\bf r}) \right],
\nonumber 
\\ && 
a_{{\downarrow}{\uparrow}}^{21}({E},{\bf k},{\bf r}) 
=\frac{i}{2} \left[ 
  f^{\dagger}({E}+i{\eta},{\bf k},{\bf r})
 -f^{\dagger}({E}-i{\eta},{\bf k},{\bf r}) \right] 
\qquad  
\label{eq:afun}
\end{eqnarray}
and $\bar{g}(E,{\bf k},{\bf r})=g(E,{\bf k},{\bf r})$. 
$T_{\rm c}$($<T_{c_0}$) is superconducting transition temperature 
at a finite magnetic field. 
We define $t=T/T_c$.
$(T_{1gg}T)^{-1}$ is the contribution in $(T_1T)^{-1}$ 
from the DOS term $W_{gg}$, and 
$(T_{1ff}T)^{-1}$ is the contribution from the coherence term $W_{ff}$.

%
\begin{figure} [thb]
\begin{center}
\includegraphics[width=8.5cm]{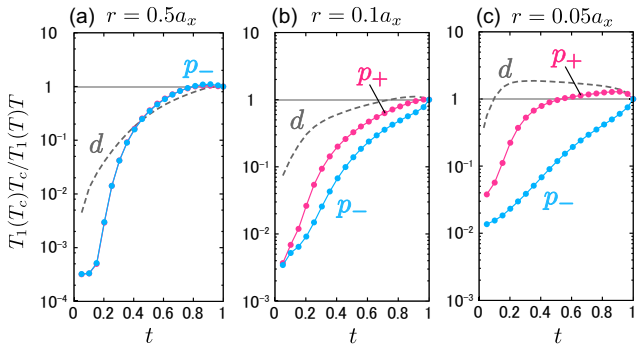}
\end{center}
\vspace{-0.7cm}
\caption{\label{fig01}
(Color online)
$T$-dependence of local $(T_1T)^{-1}$ 
for the $p_{\pm}$ states  
at radius $r/a_x = 0.5$(a), 0.1(b), 0.05({c}) from the vortex center along the NNN vortex direction.
$a_x$ is inter vortex distance along the NNN direction. 
We plot normalized values $(T_1(T)T)^{-1}/(T_1(T_c)T_c)^{-1}$ 
as a function of $t$ at $H/B_0 = 0.02$.
The vertical axis is a logarithmic scale. 
The $d_{x^2 - y^2}$-wave case is also shown for reference.
$T_c / T_{c_0} = 0.985\ (0.975)$ at $H/B_0 = 0.02$ in the $p_{\pm}$ ($d_{x^2 - y^2}$) states.  
}
\end{figure}
\begin{figure} [th]
\begin{center}
\includegraphics[width=8cm]{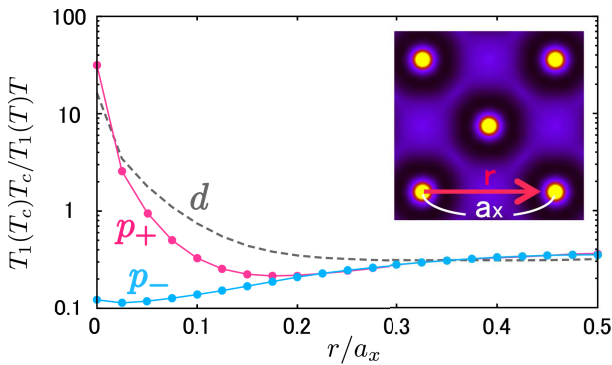}
\end{center}
\vspace{-0.7cm}
\caption{\label{fig02}
(Color online)
Local $(T_1T)^{-1}$ 
as a function of radius $r/a_x$ from the vortex center 
along the NNN direction 
for the $p_+$ and $p_-$ states. 
The $d_{x^2-y^2}$-wave case is also shown. 
The vertical axis is a logarithmic scale. 
$T/T_{c_0}=0.5$ and $H/B_0=0.02$. 
$(T_1T)^{-1}$ is normalized by the value at $T_c$.  
The inset shows a spatial structure of $(T_1T)^{-1}$ 
for the $p_+$ state. 
Brighter region has larger $(T_1T)^{-1}$.
}
\end{figure}

\section{local NMR relaxation rate}
First, we study the $T$-dependence of local $(T_1T)^{-1}$ shown 
in Fig.~\ref{fig01} for $p_{\pm}$ states.
For a reference, we also show the $d_{x^2-y^2}$-wave pairing state
$\tilde{\Delta}({\bf r},{\bf k})
=\Delta_d({\bf r})\sqrt{2}\cos 2\theta_k$~\cite{K.Tanaka2}. 
Outside of vortex core, such as the midpoint between next nearest neighbor (NNN) vortices in Fig. \ref{fig01}(a), 
the $T$-dependence is similar to the bulk chiral $p$-wave superconductors 
in both $p_\pm$ states. 
There, we see exponential $T$-dependence at low $T$, reflecting 
the full gap $|\phi_{p \pm}|=1$. 
On the other hand, around the vortex core in Figs. \ref{fig01}(b) and \ref{fig01}({c}),
we see the different behaviors 
depending on the chirality directions. 
In the $p_+$ state, $(T_1T)^{-1}$ is more enhanced with 
approaching the vortex center. 
This enhancement is due to the localized low energy DOS 
around the vortex core, and moderate compared to 
the $d_{x^2-y^2}$-wave pairing state~\cite{K.Tanaka2}. 
However, the enhancement does not occur in the $p_-$ state  
in Figs. \ref{fig01}(b) and \ref{fig01}({c}). 
The reason of this suppression is related to 
the odd-frequency Cooper pairs around the vortex core, as discussed later.

As shown in Fig. \ref{fig02}, to see the spatial dependence in detail, 
we present local $(T_1T)^{-1}$ as a function of radius $r$ 
on a line between NNN vortices. 
Outside of the vortex core $r/a_x {\ge} 0.2$,  
$(T_1T)^{-1}$ shows almost the same $r$-dependence 
between the $p_+$ and $p_-$ states. 
Inside the vortex core, it is characteristic that 
$(T_1T)^{-1}$ is enhanced in the $p_+$ state, but 
it is anomalously suppressed in the $p_-$ state.
It is also noted that $(T_1T)^{-1}$ monotonically decreases as a function of $r$ in the $d$-wave, 
but it has a minimum at $r \sim 0.175a_x$ in the $p_+$ state. 
The minimum region surrounding the vortex core is also seen 
in the spatial structure of $(T_1({\bf r})T)^{-1}$ shown 
in the inset of Fig. \ref{fig02}.

\begin{figure}[th]
\begin{center}
\includegraphics[width=6cm]{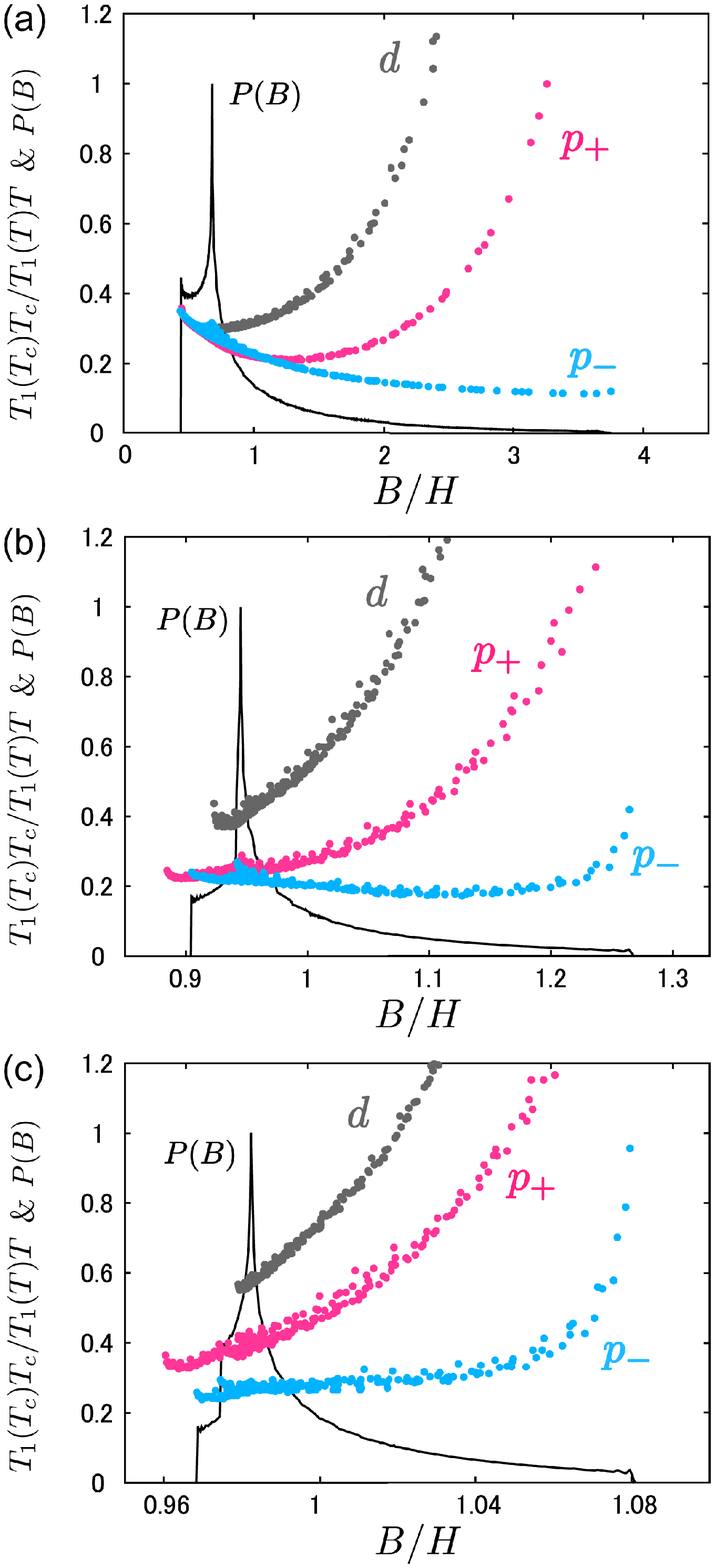}
\end{center}
\vspace{-0.7cm}
\caption{\label{fig03}
(Color online)
Solid lines indicate the Redfield pattern of 
the NMR resonance line shape, $P(B)$, for the $p_-$ state. 
Points are for $B$-dependence of $(T_1T)^{-1}$ 
for the $p_+$ and $p_-$ states. 
The $d_{x^2-y^2}$-wave case is also shown.
$T/T_{c_0} = 0.5$ and $H/B_0 = 0.02$(a), 0.10(b), 0.20({c}).  
$(T_1T)^{-1}$ is normalized by the value at $T_c$. 
Only data points $(T_1T)^{-1} {\le} 1.2$ are presented in (a), (b) and ({c}).  
}
\end{figure}

Next, we discuss how the difference between $p_+$ and $p_-$ states 
is detected in the site-selective NMR measurement. 
From the internal field distribution 
${\bf B}({\bf r})$, 
we theoretically obtain the Redfield pattern~\cite{Redfield} 
of the NMR resonance line shape, as 
$P(\omega) = {\int} {\delta}(\omega - B({\bf{r}})) d{\bf{r}}$,  
since the intensity at each resonance frequency $\omega$ comes 
from the volume satisfying
${\omega} = B({\bf{r}})$ in a unit cell.
In ${\rm Sr_2RuO_4}$, $P(B)$ was observed by ${\rm \mu SR}$~\cite{Aegerter}. 
In Fig.~\ref{fig03}(a), with $P(B)$,  
we plot local $(T_1T)^{-1}$ as a function of 
local field $B({\bf{r}})$ at the same position ${\bf r}$. 
At lower resonance fields $B/H <1$ near the peak of $P(B)$, 
NMR signals come from outside of the vortex cores. 
In this range, $(T_1 T)^{-1}$ decreases as a function of $B$ 
in both $p_{\pm}$ states similarly. 
The tail of $P(B)$ at higher $B$ is approaching the vortex center. 
In this range $B/H >1$, we can see the chirality dependence, 
i.e., $(T_1T)^{-1}$ increases as a function of $B$ in the $p_+$ state, 
but it decreases in the $p_-$ state. 
However, at the low applied field $H$, the signal of the vortex core contribution at higher $B$ 
is weak in $P(B)$. 
On the other hand,
at higher applied field $H$ as shown in Figs. \ref{fig03}(b) and  \ref{fig03}({c}), 
the signal for distinguishing the chirality becomes larger in $P(B)$,  
since weight of the vortex core region increases 
within the unit cell of the vortex lattice with increasing $H$. 
In Fig. \ref{fig03}(b), $(T_1T)^{-1}$ in the $p_+$ state increases as a function of $B$ in all resonance filed range,
while it is almost flat in the $p_-$ state except for largest $B$.
In Fig. \ref{fig03}({c}), $(T_1T)^{-1}$ in both $p_{\pm}$ states increases as a function of $B$.
From these calculation results,
we can identify the direction of the chirality i.e., $p_+$ or $p_-$ state
by measuring $B$-dependence of $(T_1T)^{-1}$.
In particular, 
it is important that we observe the monotonically decreasing or flat behavior of $(T_1T)^{-1}$ as a function of $B$,
since this behavior is realized only in the $p_-$ state.
And, from the previous studies~\cite{GL-chiral, Ichioka-chiral}, 
it is expected that the $p_-$ state has a lower free energy than the metastable $p_+$ state in the vortex state.
However, we should be careful about strength of applied field,
since $B$-dependence of $(T_1T)^{-1}$ changes as shown in Fig. \ref{fig03}({c}), 
when the applied field is too high.

\begin{figure} 
\begin{center}
\includegraphics[width=6cm]{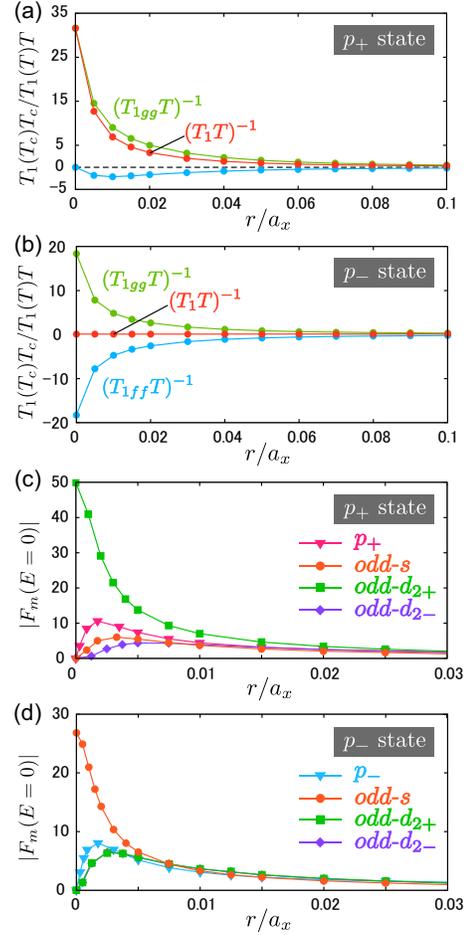}
\end{center}
\vspace{-0.7cm}
\caption{\label{fig04}
(Color online)
$r$-dependence of $(T_1T)^{-1}$, $(T_{1gg}T)^{-1}$, $(T_{1ff}T)^{-1}$ 
in (a) the $p_+$ state and (b) the $p_-$ state. 
$(T_1T)^{-1}$, $(T_{1gg}T)^{-1}$, $(T_{1ff}T)^{-1}$ is normalized by $(T_1(T_c)T_c)^{-1}$.
$r$-dependence of orbital-decomposed Cooper pair's amplitude 
$|{\mathcal{F}}_m (E=0)|$ 
in (c) the $p_+$ state and (d) the $p_-$ state. 
$m=s$, $p_{\pm}$, and $d_{2 \pm}$. 
In all figures, $T/T_{c_0}=0.5$ and $H/B_0=0.02$. 
$r$ is radius from the vortex center along the NNN vortex direction.
In the $p_-$ state, 
$|{\mathcal{F}}_{d_{2+}} (r, E=0)| {\sim} |{\mathcal{F}}_{d_{2-}} (r, E=0)|$.
}
\end{figure}

\begin{table}
\begin{tabular} {|c|c|l|l|} \hline
Symmetry  & Chirality & \multicolumn{2}{c|}{Vorticity $W$ }
  \\ \cline{3-4}
component & $L_z$     & $p_+$ state & $p_-$ state   \\ \hline 
$d_{2+}$ & 2  & \ 0 (center) & \ $-2$ \\ \hline 
$p_{+}$  & 1  & \ 1 (main) & \ $-1$ \\ \hline 
$s$      & 0  & \ 2  &  \ 0 (center) \\ \hline 
$p_{-}$  & $-1$  & \ 3  & \ 1 (main) \\ \hline 
$d_{2-}$ & $-2$  & \ 4  & \ 2 \\ \hline 
        &     & $L_z+W=2$  & $L_z+W=0$ \\ \hline 
\end{tabular}
\caption{\label{table1}
Relation of vorticity $W$ and chirality $L_z$ for each symmetry component 
of the orbital-decomposed Cooper pair ${\mathcal{F}}_m$ around a vortex 
in the $p_+$ and $p_-$ states. 
Main component in each case has $W=1$. 
The induced component has other $W$ locally around the vortex center
by the conservation of $L_z + W$~\cite{Tanuma}.
At the vortex center, induced component with $W=0$ has finite amplitude.  
}
\end{table}

\section{Relation to Odd-frequency Cooper pairs}
To discuss the reasons for the anomalous suppression of $T_1^{-1}$
around the vortex core in the chiral $p$-wave superconductors,
we present the decomposition of $(T_1T)^{-1}$ to 
the DOS term  $(T_{1gg}T)^{-1}$ and the coherence term $(T_{1ff}T)^{-1}$
in Figs. \ref{fig04}(a) and \ref{fig04}(b). 
There, we see that $(T_{1gg}T)^{-1}$ is enhanced around the vortex core 
in both $p_{\pm}$ states similarly, 
as in the $s$- and $d_{x^2-y^2}$-wave cases~\cite{K.Tanaka2}. 
The enhancement reflects low energy DOS around the vortex core. 
The chirality-dependence appears in negative coherence term $(T_{1ff}T)^{-1}$. 
In the $p_-$ state, negative $(T_{1ff}T)^{-1}$ cancels 
the enhancement of $(T_{1gg}T)^{-1}$, 
so that $(T_{1}T)^{-1}$ is suppressed in the vortex core. 
In the $p_+$ state, weak suppression of $(T_{1}T)^{-1}$ 
in the region surrounding vortex in Fig. \ref{fig02} is also 
due to the small negative term $(T_{1ff}T)^{-1}$. 
Therefore, in the $p_+$ state, 
we can say that the $(T_{1gg}T)^{-1}$ of a normal signal obscures 
the $(T_{1ff}T)^{-1}$ of a superfluid response.
However, in the $p_-$ state, 
since the superfluid response is enhanced around the vortex core 
including the proximity effect of superconductivity, 
the normal signal does not obscure
the superfluid response.

At last, 
we discuss origin of the negative coherence term.
From Eqs. (\ref{eq:T1})-(\ref{eq:afun}),
$s$-wave pair can contribute to the coherence term $(T_{1ff}T)^{-1}$ 
since the condition ${\langle}f{\rangle}_{\bf{k}} {\ne} 0$ with $L_z=0$. 
Actually,
in conventional $s$-wave superconductor, 
a Hebel-Slichter peak appears below $T_c$ 
due to the coherence term~\cite{Hebel-Slichter, Masuda, K.Tanaka2}.  
To check this condition, we calculate orbital-decomposed Cooper pair 
${\mathcal{F}}_m (E,{\bf r})
= {\langle} {\phi}_m^\ast({\bf k})
  f(E+{\rm i}\eta,{\bf{k}},{\bf r}){\rangle}_{\bf{k}}$.
In addition to $\phi_{p \pm}({\bf k})$, 
we employ $\phi_{s}({\bf k})=1$ for the $s$-wave, and 
$\phi_{d2 \pm}({\bf k})={\rm e}^{\pm i 2 \theta}$ for the chiral $d$-wave. 
The obtained $s$- and $d$-wave components in the chiral $p$-wave superconductors 
are odd-frequency Cooper pair~\cite{Tanuma}.
In Figs. \ref{fig04}(c) and  \ref{fig04}(d), 
we present the $r$-dependence of $|{\mathcal{F}}_m (E=0,{\bf r})|$, 
where the induced $s$- and $d$-wave amplitude have large values 
around the vortex core. 
As summarized in Table \ref{table1}, 
the vorticity $W$ of the symmetry component ${\mathcal{F}}_m$ 
with the chirality $L_z$ 
is determined by the condition $L_z+W=2$ in the $p_+$ state and 
$L_z+W=0$ in the $p_-$ state. 
In the $p_+$ state, chiral $d_{2+}$-wave component has $W=0$, 
giving large amplitude at the vortex center. 
Small induced $s$-wave component also appears, 
but it vanishes at the vortex center since it has $W=2$, as shown in Fig. \ref{fig04}(c).
In the $p_-$ state, 
the $s$-wave component has $W=0$ thus it has large amplitude 
at the vortex center, as shown in Fig. \ref{fig04}(d).
The odd-frequency $s$-wave Cooper pair 
determines the $r$-dependence of the negative coherence term 
$(T_{1ff}T)^{-1}$ in Figs. \ref{fig04}(a) and \ref{fig04}(b).
In particular, at low $T$ limit, we confirmed that $(T_{1ff}T)^{-1} {\sim} -|{\mathcal{F}}_s (E=0)|^2$ at the vortex center
from the calculation results.

The previous theoretical study using the Andreev bound state model showed that
$T_1^{-1}$ at the vortex center is completely zero $(T_1^{-1} {\sim} 0)$ due to the coherence effect
when the $L_z$ is anti-parallel to the $W$~\cite{Kato-NMR}.   
On the other hand,
previous our study using the Bogoliubov-de Gennes theory confirmed the relation 
$N(E=0, {\bf r}) {\propto} |{\mathcal{F}}_s (E=0,{\bf r})|$ in the $p_-$ state 
for the vortex core quasiparticle states with Majorana zero mode~\cite{Daino}.
Considering these relation, we find that the $(T_{1ff}T)^{-1}$ related to the odd-frequency $s$-wave Cooper pair 
tends to cancel the local DOS term $(T_{1gg}T)^{-1}$,
since $(T_{1gg}({\bf{r}})T)^{-1} {\sim} N(E=0, {\bf r})^2$ and $(T_{1ff}({\bf{r}})T)^{-1} {\sim} -|{\mathcal{F}}_s (E=0,{\bf r})|^2$
at low $T$ and $H$ limit (low energy limit). 
Therefore, the anomalous suppression of $(T_1T)^{-1}$ is also
explained by the nature of Majorana state.
Note that,  
in our calculation results at finite $T$ and $H$ states,  
$(T_1T)^{-1}$ is not completely zero around the vortex core,
since quasiparticle states different from Majorana zero mode also contribute to the NMR relaxation,
as shown in Fig.~\ref{fig02}.

When we discuss the influence of the sub-dominant components,
we have to distinguish the order parameter ${\Delta}$ and the pair amplitude ${\mathcal{F}}$.
The sub-dominant components such as odd-frequency $s$- and $d$-wave Cooper pairs vanish in the order parameter, 
since the order parameter is determined by the gap equation of Eq. (\ref{eq:scD}).
Therefore, the qualitatively unique mechanism of negative coherence term related to the odd-frequency Cooper pairs 
in the chiral $p$-wave superconductors does not seriously depend on the details of 
setting the pairing interaction for the sub-dominant order parameter. 

\section{Summary}
We have calculated the $T$-, $r$- and $B$-dependence
of the local NMR relaxation rate $(T_1T)^{-1}$ 
in two chiral $p_{\pm}$ states, and $d_{x^2-y^2}$-wave as a reference.
We have clarified that $(T_1T)^{-1}$ in the $p_+$ state is enhanced with approaching the vortex center 
by the contribution of low energy excitations of the vortex core,
but it is anomalously suppressed around the vortex core in the $p_-$ state.
This chirality-dependence of local $(T_1T)^{-1}$ may be observed
by the site-selective NMR measurement via the $B$-dependence of $(T_1T)^{-1}$ in $P(B)$.
Further, we have theoretically found that the anomalous suppression of $(T_1T)^{-1}$  
around the vortex core is  
due to the negative coherence term 
by the induced odd-frequency $s$-wave Cooper pair with Majorana state. 

We hope that these theoretical estimates of local $(T_1T)^{-1}$ will be confirmed by the site-selective NMR measurement,
and will be used for detecting the pairing symmetry with chirality in the chiral $p$-wave superconductors,
and natures of odd-frequency Cooper pairs and Majorana states.


\end{document}